\documentstyle[aps,prb,epsfig,floats]{revtex}
\begin{document}
\draft \wideabs{
\title{Fiber-coupled Antennas for Ultrafast Coherent Terahertz Spectroscopy
in Low Temperatures and High Magnetic Fields}
\author{S. A. Crooker}
\address{National High Magnetic Field Laboratory - LANL, MS E536, Los Alamos, NM 87545}
\date{27 March 2002}
\maketitle
\begin{abstract}
For the purposes of measuring the high-frequency complex
conductivity of correlated-electron materials at low temperatures
and high magnetic fields, a method is introduced for performing
coherent time-domain terahertz spectroscopy directly in the
cryogenic bore of existing dc and pulsed magnets.  Miniature
fiber-coupled THz emitters and receivers are constructed and are
demonstrated to work down to 1.5 Kelvin and up to 17 Tesla, for
eventual use in higher-field magnets.  Maintaining the sub-micron
alignment between fiber and antenna during thermal cycling,
obtaining ultrafast (${<200}$~fs) optical gating pulses at the end
of long optical fibers, and designing highly efficient devices
that work well with low-power optical gating pulses constitute the
major technical challenges of this project.  Data on a YBCO
superconducting thin film and a high mobility 2D electron gas is
shown.

\end{abstract}
\pacs{PACS numbers:}
} \narrowtext {\bf Introduction}

Time-domain terahertz spectroscopy is an established technique for
the measurement of high-frequency conductivity, typically in the
range between 100 GHz and ${\sim}$3000 GHz. Pioneered in the late
1980s\cite{Smith,Fattinger}, ``table-top" terahertz spectrometers
employing photoconductive antennas gated by ultrafast optical
pulses have been used to study a wide range of material systems,
including semiconductors and dielectrics\cite{Grisch}, normal and
high-$T_c$ superconductors\cite{Nuss1,Nuss2},
liquids\cite{Pedersen}, flames\cite{Cheville}, and
gases\cite{Harde}. This THz frequency range lies between that
which is readily accessible by microwave cavity techniques (on the
low frequency side), and Fourier-transform infrared spectroscopies
(on the high frequency side).  These frequencies correspond to
energies between 0.4 meV and ${\sim}$12 meV, or alternatively,
temperatures between 4K and 140K and magnetic fields between 4
Tesla and 100 Tesla.

This is precisely the energy, temperature, and in particular the
magnetic field scale relevant to many novel correlated-electron
systems of interest today, including high-$T_c$ superconductors
(where the upper critical field ${H_{c2}}$ corresponds to tens or
even hundreds of Tesla\cite{Boebinger}), heavy-fermion and
Kondo-insulating materials (where, {\it e.g.}, the Kondo
spin/charge gap in Ce$_3$Bi$_4$Pt$_3$ may be closed above 30
T\cite{Jaime}), colossal magnetoresistive manganites (melting of
charge/orbital order at high fields\cite{Tokura}), 2D electron
gases (composite fermion dynamics in the high-field fractional
quantum Hall regime), and organic metals (novel field-{\it
induced} superconductivity above 17~T\cite{Uji}). Thus it is of
keen interest to perform measurements of the complex THz
conductivity not only in the regime of low temperatures, but also
at high magnetic fields. However, the conventional ``table-top"
transmission terahertz spectrometer is a rather involved and
physically large setup, typically utilizing several
micropositioning stages to align the THz antennas with respect to
the free-space laser beams, and off-axis parabolic optics to
collimate and focus the terahertz pulses over short distances.
These traditional methods work extremely well, but are not
compatible with high-field magnets (10-60~T), which are generally
solenoids with narrow, cryogenic bores accessible primarily via
meters-long experimental probes.

To this end we have developed extremely sensitive, miniaturized,
optical fiber-coupled THz emitters and receivers for remote use
directly in the low-temperature bore of a high-field (dc or
pulsed) magnet.  These devices permit ultrafast, coherent,
time-domain THz transmission spectroscopy of samples in the
frequency range between 100 GHz and ${\sim}$2000 GHz. Due to the
coherent nature of the detection (both amplitude and phase of the
THz electric field are measured), the complex conductivity may be
directly evaluated without the need for Kramers-Kronig analyses,
in contrast with phase-incoherent schemes based on thermal
cyclotron emission\cite{Burke}. The gated nature of the detection
permits high signal-to-noise data with minimal (${\sim}$2 mW)
optical power input and, where necessary, the ability to acquire
complete spectra in tens of milliseconds. The primary challenges
of this project include maintaining sub-micron alignment between
fiber and antenna upon repeated thermal cycling, achieving
ultrafast (${<200}$~fs) optical pulses at the end of tens of
meters of singlemode optical fiber, and obtaining complete
time-domain scans with high signal-to-noise ratio using only
milliwatts of optical power, no lock-in detection, and (for pulsed
magnets) only ${\sim}$100 ms of integration time.

\bigskip
{\bf Experimental setup}

Using photoconductive antennas, wide-bandwidth THz pulses may only
be generated (and detected) by gating the antennas with ultrafast
optical pulses, and for this reason it is necessary to compensate
for the positive group-velocity dispersion (GVD) of optical fibers
so that fast optical pulses may be obtained at the ends of long
fibers.  Normal silica optical fiber exhibits a GVD of roughly 120
fs/m-nm at 800 nm, so that without compensation, a 100 fs input
optical pulse with a bandwidth of 10~nm broadens, in the best
case, to ${>20}$ picoseconds after a typical 20 meter length of
fiber. Such a lengthy optical pulse is useless for generating or
detecting THz radiation. Thus it is necessary to precompensate and
impose a negative chirp on the optical pulses before launching
into the optical fibers, so that the optical pulses shorten in
time as they travel through the fiber and achieve a minimum value
right at the THz devices.

\begin{figure}
\epsfig{file=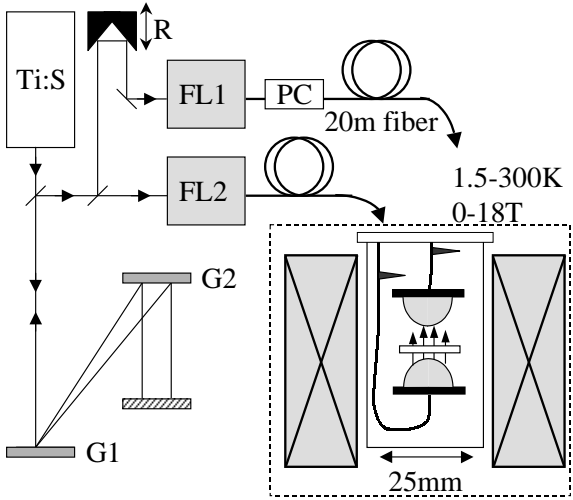, width=3.1in,clip=}
\vspace{0.1in} \caption{Experimental schematic. Ultrafast optical
pulses are pre-chirped (stretched) by gratings G1 and G2, and
coupled into fibers by fiber launchers FL1 and FL2. A
rapid-scanning retroreflector (R) and polarization controller (PC)
are also present in the pulse train for the THz receiver. Pulses
achieve minimum temporal width (and highest peak intensity) at the
photoconductive THz emitter and receiver, which are located in the
cryogenic bore of a high-field magnet.} \label{fig1}
\end{figure}

The experimental schematic for THz transmission spectroscopy in
high-field magnets is illustrated in Fig. 1.  Ultrafast optical
pulses (100~fs, centered at 800~nm)  from a commercial Ti:sapphire
laser are directed to a two-grating pulse stretcher, which imparts
a negative chirp (blue wavelengths leading red wavelengths) onto
the pulses. The magnitude of the negative chirp, tuned via the
separation between the two gratings G1 and G2 (1200 grooves/mm),
is chosen to optimally compensate for the intrinsic positive GVD
of optical fiber, and thus depends on the laser wavelength and the
length of fiber used.  For typical fiber lengths of 20 meters, the
100 fs, transform-limited pulses are stretched to  approximately
24 picoseconds. After leaving the pulse stretcher, the
negatively-chirped pulse train is equally split into an ``emitter"
and a ``receiver" beam.  The former is launched directly into a
singlemode optical fiber, while the latter is delayed by a
scanning retroreflector before being launched into another
singlemode fiber of equal length.  As the optical pulses travel
from the laboratory to the low-temperature probe housing the THz
antennas, their temporal width decreases due to the positive GVD
of the fiber, achieving a minimum pulsewidth directly at the
lithographically-defined photoconductive antennas. The minimum
pulsewidth depends critically on pulse energy, fiber length, and
stretcher position as discussed in detail below.   The receiver's
fiber passes through a polarization controller (PC), as it was
found that the signal amplitude depends slightly (${\pm 15\%}$) on
the polarization of the optical pulses incident on the THz
receiver.

\begin{figure}
\epsfig{file=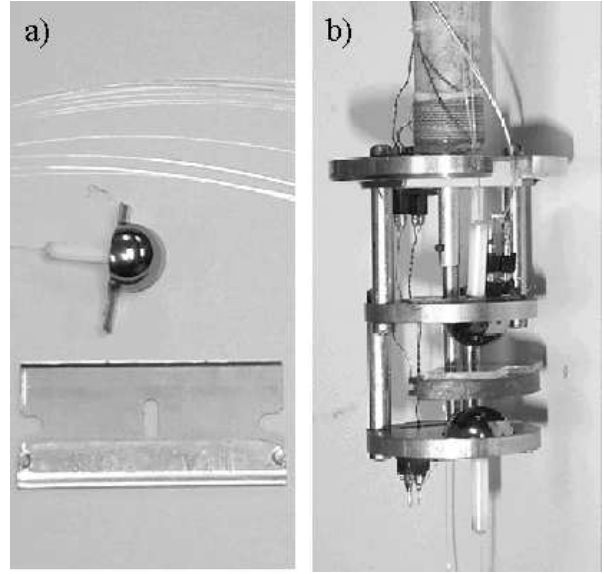, width=3.1in,clip=}

\vspace{0.1 in} \caption{a) Photographs of a) a fiber-coupled THz
antenna (next to a standard razor blade), and b) the lowest
section of the cryogenic high-field THz probe.} \label{fig2}
\end{figure}

In an experiment, the sample of interest is positioned between the
THz emitter and receiver in the cryogenic bore of the magnet.
Following the usual prescription, an ultrafast optical pulse
promotes mobile carriers in the biased stripline emitter, and the
subsequent current surge generates a burst of broadband THz
radiation which is coupled into free space through a silicon
hyperhemispherical substrate lens of radius 2-5 mm. The height of
the substrate lens is chosen to collimate the THz radiation. After
passing through the sample, the THz pulse is focussed onto the
stripline receiver by another substrate lens, where the
``instantaneous" THz electric field is gated by a second ultrafast
optical pulse, generating a measurable current. The complete
time-dependent THz electric field is mapped by rapidly scanning
the time delay between the excitation and gating optical pulses,
and the amplified current signal is sent directly to a digitizing
oscilloscope. Photographs of a fiber-coupled antenna and of the
actual apparatus are shown in Figure 2. The THz emitter and
receiver are mounted facing one another with a gap of roughly 1
cm.  The emitter bias is provided by twisted-pair leads and
external batteries (10-20 V, typically).  The detected
photocurrent is coupled from the receiver to an external current
amplifier (DL Instruments model 1212) via a micro-coaxial cable
selected for its low loss and insensitivity to microphonic noise.
A rotating copper sample stage enables the sample to be moved out
of the THz beam path, so that a reference scan (crucial for
quantitative interpretation of data) may be taken at each new
temperature or field.  Temperature control is provided by a Cernox
thermometer in the sample stage and a wire heater on the body of
the probe.  The entire probe is 25 mm in diameter, and may be used
in vacuum, vapor, or liquid helium environments .

\bigskip
{\bf Pulse dispersion management in long fibers}

The power and bandwidth of the generated THz radiation is strongly
influenced by the temporal width of the ultrafast optical pulses
that are used to gate the antennas.  Fig. 3 shows the signal
(${\propto}$ THz electric field) for a fiber-coupled
emitter/receiver pair that are driven by 800~nm optical pulses of
constant energy (26 pJ/pulse), but varying pulsewidth.  Here, the
optical pulsewidth at the antennas is varied between 175 fs and
1300 fs by changing the spacing between the gratings in the pulse
stretcher. The effects are immediately clear: Both the amplitude
and bandwidth of the measured THz radiation decrease as the
temporal width of the optical gating pulses is increased from the
minimum value of 175 fs.  The drop in power is particularly marked
at higher frequencies; {\it e.g.}, above 800 GHz the power
spectrum of the THz pulses (Fig. 3b) decreases over an order of
magnitude.

After propagating through a typical length of 10-20~m of optical
fiber, the minimum temporal width of an optical pulse is always
somewhat larger than the original 100 fs, even when the positive
GVD of the optical fiber is optimally balanced by the pulse
stretcher.  The main reasons for this are uncompensated cubic
phase dispersion in the stretcher/fiber system, and self-phase
modulation (SPM) in the optical fiber.  Quadratic, cubic, and
higher-order phase dispersion of an optical pulse in fiber arises
simply because the index of refraction of silica (or any material)
is not constant with wavelength, and thus different wavelengths
within a broadband ultrafast optical pulse travel with different
group velocities. Whereas the negative quadratic phase dispersion
(GVD) imparted by the pulse stretcher may be tuned to exactly
cancel the positive quadratic phase dispersion of the fiber, both
the stretcher and the fiber impart a small cubic phase dispersion
onto the optical pulse, and these contributions do not cancel.
Indeed, it is only by rather involved design, or via the inclusion
of phase-shifting optics, that cubic phase dispersion may be
compensated\cite{Shen}. For the purposes of generating and
detecting THz radiation, the pulse broadening due to cubic phase
dispersion does not degrade the signal significantly to warrant
correction, although for longer lengths of fiber (${>50}$~m)
correction may be necessary.

Self-phase modulation, on the other hand, is an inherently
nonlinear optical phenomenon arising from an {\it
intensity-dependent} phase shift, whose effect is to redistribute
energy within an optical pulse to different spectral components.
Thus, unlike the linear effects of GVD (quadratic) or cubic phase
dispersion, SPM acts to create or remove additional wavelengths
within the optical pulse. Because SPM in fibers arises from the
high instantaneous power in ultrafast optical pulses, it may be
minimized by using very low-power pulses, typically of order 30 pJ
or less for 100 fs pulses.

\begin{figure}
\epsfig{file=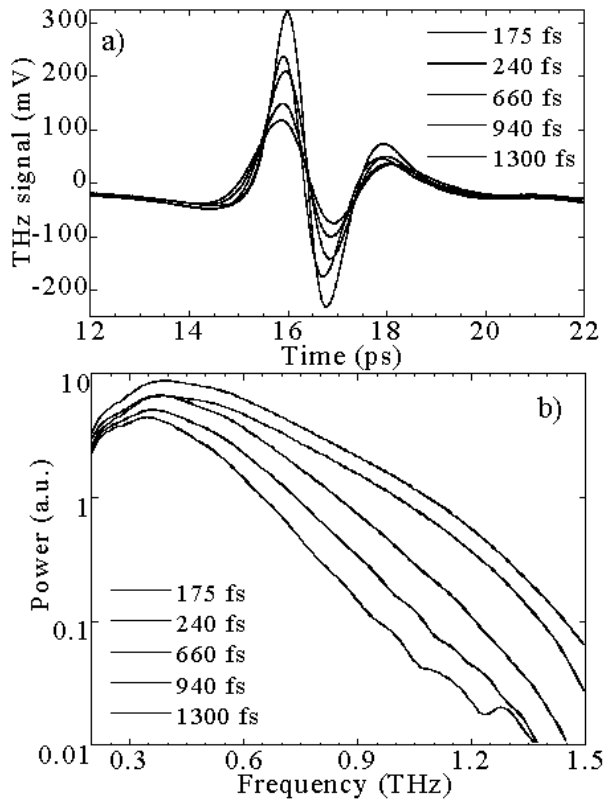, width=3.1in,clip=}
\vspace{0.1in} \caption{The measured ultrafast THz electric field
from a fiber-coupled emitter/receiver pair as a function of the
temporal width of the 800 nm optical gating pulses. The fiber
length is 10m b) The corresponding power spectrum.} \label{fig3}
\end{figure}

The combined effects of SPM, cubic phase dispersion, pulse power,
and stretcher position on the optical pulsewidth are shown in
Figure 4.  Here, prechirped 100 fs low-power (closed circles) and
high-power (open circles) optical pulses centered at 800 nm are
launched into 20 meters of optical fiber, and the temporal width
of the pulses at the end of the fiber is measured as a function of
stretcher position. The pulsewidth is measured via autocorrelation
and a ${sech^2(t)}$ pulse shape is assumed.  At one extreme, when
the distance between the two stretcher gratings is 15 mm longer
than the optimum separation of 160 mm, the pulses are
``overstretched", and therefore never achieve their minimum value
in the optical fiber ({\it i.e.}, the pulses are still somewhat
negatively chirped at the fiber exit).  As the distance between
the gratings is decreased, the pulses exit the fiber with shorter
and shorter pulsewidth until an optimum position is reached and
the GVD of the fiber is exactly compensated. Further decreasing
the distance between the gratings results in ``understretched"
pulses, where the minimum pulsewidth is achieved somewhere within
the optical fiber (rather than at the end).  Note that the minimum
pulsewidth for the low-power pulses (250 fs) is larger than the
100 fs input pulsewidth. Since further reduction of the pulse
energy has very little effect on the shape of this curve (not
shown), the 250 fs pulsewidth is due to the uncompensated cubic
phase dispersion of the stretcher-fiber combination (also, similar
experiments in 10, 5, and 1 meter fibers show that the  minimum
pulsewidth approaches 100 fs).

\begin{figure}
\epsfig{file=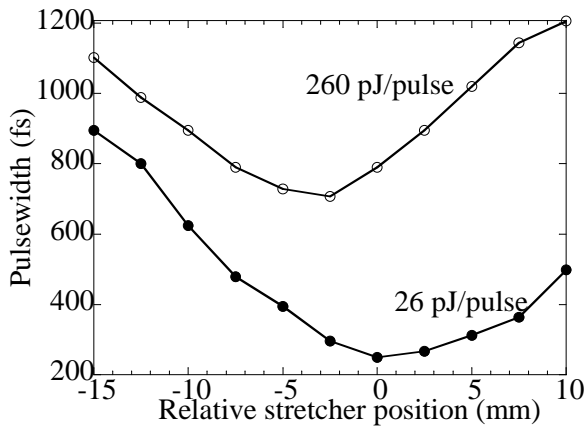, width=3.1in,clip=}
\vspace{0.1in} \caption{The temporal width of the optical gating
pulses, measured at the end of 20 m of singlemode fiber, as the
distance between gratings G1 and G2 is varied.   Solid (open) dots
correspond to low (high) power pulses.  Lines are guides to the
eye.} \label{fig4}
\end{figure}

More interesting are the data from the high-power pulses, which
exhibit a much larger minimum pulsewidth of 700 fs (and, this
minimum value occurs at a slightly different position of the
stretcher). In this case, the effects of nonlinear SPM, arising
from the large instantaneous intensity of the high-power pulse,
actually {\it narrows} the spectral bandwidth of the optical
pulse, thereby preventing the optical pulse from attaining a small
temporal width, even if the pulse were transform-limited.
SPM-induced spectral narrowing in a 10 meter optical fiber is
explicitly shown in Fig. 5.  As a function of pulse energy, the
minimum attainable pulsewidth at the fiber exit is shown in Fig.
5a.  Figure 5b shows the associated spectral content of these
pulses, normalized for comparison.  For pulse energies below 26
pJ/pulse, the minimum obtainable pulsewidth remains unchanged at
${\sim}$175 fs, and the spectral content of these pulses is
relatively constant and equal to the spectrum of the pulses before
entering the optical fiber, indicating that nonlinear (SPM)
effects in the fiber are not significant. For pulse energies above
26 pJ, SPM causes the minimum pulsewidth to grow dramatically from
175 fs, reaching a value of 400 fs at 260~pJ/pulse, or 20~mW of
average laser power. The spectral bandwidth of the pulses shrinks
correspondingly, from a value of 10 nm for low-power pulses, to
just under 3.5~nm for the 260 pJ pulses. Although it stands in
contrast to the much more well-known and widely exploited spectral
{\it broadening} which occurs for positively-chirped or
transform-limited optical pulses, SPM-induced spectral narrowing
of negatively-chirped optical pulses has been previously
studied\cite{Washburn}, and further, might even be exploited as
part of a system for fiber delivery of ultrafast {\it nano}joule
optical pulses\cite{Clark}. Regardless, it is clear that the
deleterious effects of SPM on the temporal width of the optical
pulses (and subsequently on the THz bandwidth and power) requires
that the optical pulse energies be kept very low -- less than
${\sim}$30 pJ/pulse, or 2.3 mW average power from the 76 MHz
repetition rate Ti:S laser.

\begin{figure}
\epsfig{file=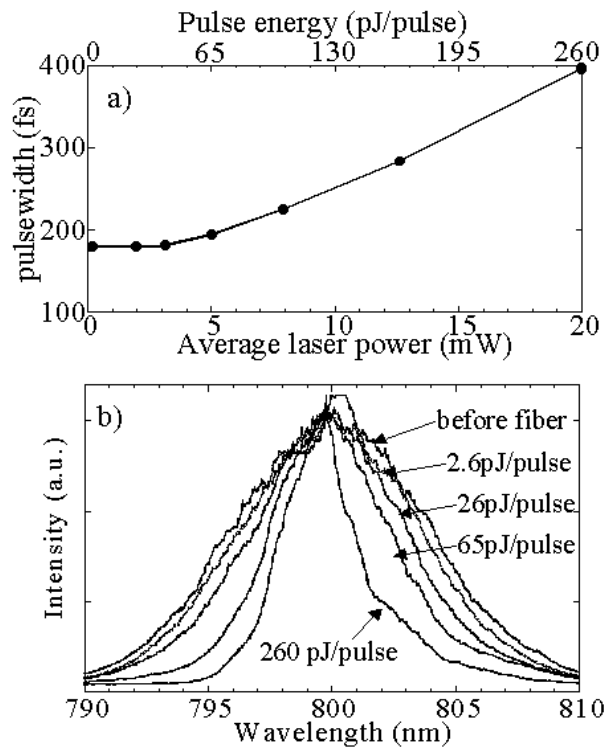, width=3.1in,clip=}
\vspace{0.1in} \caption{a) The minimum temporal width of the
optical gating pulses, measured at the end of 10 meters of
singlemode fiber as a function of pulse power. b) The
corresponding spectrum of the pulses at the end of the fiber. The
increased temporal width and spectral narrowing above 26 pJ/pulse
are due to effects of nonlinear self-phase-modulation.}
\label{fig5}
\end{figure}

\bigskip
{\bf Antenna design and construction}

The stringent upper limit on the energy of the optical gating
pulses mandate the use of specially-designed THz emitter and
receiver antennas for maximum emission efficiency and detection
sensitivity when coupled to single-mode optical fibers.
Additionally, to minimize the thermal load at liquid helium
temperatures (${<4}$ K), it is desirable to design THz antennas
which operate efficiently using as little laser power as possible
(in practice, 1-2 mW). For the most part, the THz devices are
``standard" photoconductive striplines\cite{Fattinger} deposited
via photolithography onto semiconductor substrates, with 10-20
micron wide, 1 cm long, titanium-gold lines separated by 50-100
microns. Following the work of Brener {\it et al.}\cite{Brener},
the THz emitter also incorporates opposing triangular features at
the midpoint of the stripline to concentrate the electric field in
a localized region so as to enhance the THz output. The gap
between the tips of the triangular features is 5 microns, chosen
to match the mode-field diameter of the single-mode fiber.
Similarly, the THz stripline receiver incorporates a standard
dipole, again with a 5 micron gap. The semiconductor substrates
are ErAs/GaAs superlattices grown by molecular-beam epitaxy on
GaAs wafers\cite{Kadow}. Here, the ErAs material traps carriers
efficiently, resulting in the desired subpicosecond carrier
lifetimes necessary for THz photomixing applications.
Functionally, this material system is related to
low-temperature-grown GaAs (LT-GaAs), but with the added
advantages of greater flexibility and control over sample
parameters.  THz receivers made on ErAs/GaAs superlattices were
found to exhibit excellent sensitivity to THz electric fields, on
par with that of LT-GaAs and over an order of magnitude more
sensitive (per unit power of the optical gating pulse) than
implanted silicon-on-sapphire substrates.

Good alignment and bonding of the single-mode optical fiber to the
small ``sweet spot" of the THz antenna is critical to the
construction of a usable THz device, as well over half of the THz
signal may be lost if the fiber and antenna become misaligned by
as little as 1 micron.  Because the devices are used in cryogenic
environments and must undergo repeated thermal cycling, proper
selection of materials and careful construction are necessary to
minimize differential thermal contraction which leads to
misalignment (or destruction) of the device.  Fiber-coupled THz
devices are constructed in the following way:  Using an infrared
microscope, high-resistivity silicon hyperhemispheres are centered
exactly over the antenna dipole on the back side of the wafer, and
bonded with a dilute mixture of rubber cement in toluene.
Single-mode optical fibers are then epoxied into ceramic ferrules,
and polished flush with the face of the ferrule. The ceramic
ferrule is then positioned over the THz antenna using an XYZ
translation stage with sub-micron actuators (Thorlabs MDT611). The
THz emitter is biased (10-20 V), the THz receiver is connected to
a fast current amplifier, and optical pulses are coupled into the
fibers so that the THz signal may be monitored in real time and
used to determine the optimum position of the fiber/ferrule with
respect to the antenna. Maximum signal is achieved when the
ferrule is nearly in contact with and is positioned exactly over
the antenna dipole. Bonding of the ferrule to the antenna is
achieved by backing the ferrule away from the antenna and applying
an extremely thin layer of clear, degassed Stycast 1266 epoxy to
the end of the ferrule, after which the ferrule may again be
brought into contact with the antenna and the signal maximized.
Once optimized, the epoxy may be left to cure. During alignment,
using extremely low-power optical pulses and using viscous
(slightly cured) epoxy avoids the formation of bubbles due to
local heating from the optical pulses. The resulting bond formed
between the ferrule and the antenna is quite robust and rigid, and
most importantly, does not suffer any misalignment upon repeated
thermal cycling from room temperature to cryogenic temperatures.
Use of the ceramic ferrule is important to provide mechanical
support for the optical fiber, to match the thermal contraction of
the GaAs substrate, and to facilitate the use of as little epoxy
as possible. Most epoxies, being polymer-based, exhibit thermal
contraction that is much larger than the contraction of the GaAs
substrate upon which the THz antenna resides (1.1\% total
contraction from 300K to 4K for Stycast 1266, as compared to
~${\sim}$0.1\% net contraction for GaAs), so that thick layers of
epoxy invariably shatter the substrate upon cooldown.  Devices
made with stainless steel ferrules occasionally fail upon
cooldown, presumably due to the increased mismatch of thermal
contraction between stainless steel (0.3\% net contraction) and
GaAs. Standard UV-curing optical cements  performed poorly under
cryogenic conditions, and sub-micron alignment was difficult to
maintain during the curing process. To make fiber-coupled THz
devices for room-temperature use only, any combination of ferrule
and epoxy works well -- even gluing a bare cleaved fiber directly
to the antenna with a drop of Stycast 1266 is quite reliable, if
somewhat flimsy. Alternatively, for room-temperature antennas
coupled by short lengths of fiber, at least one very nice complete
commercial system exists for THz spectroscopy and imaging
applications\cite{Rudd}.

\begin{figure}
\epsfig{file=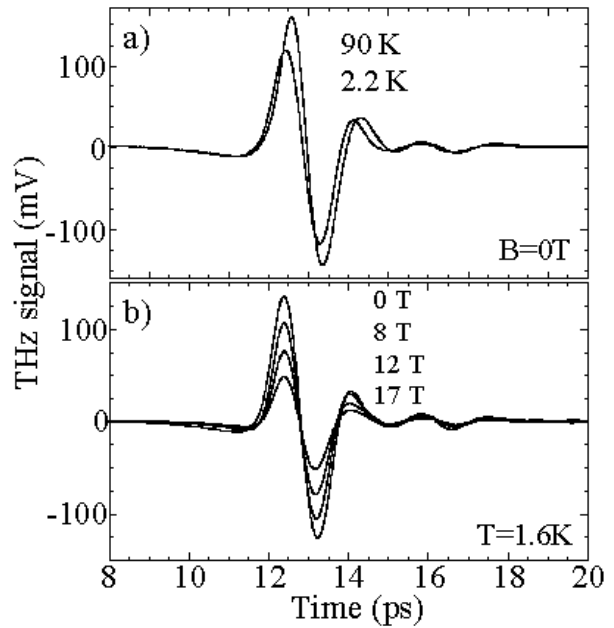, width=3.1in,clip=}
\vspace{0.1in} \caption{a) The measured THz electric field from an
in-situ fiber-coupled emitter/receiver pair at cryogenic
temperatures, demonstrating a weak temperature dependence and
phase shift.  b) Same, but as a function of  magnetic field at
1.6K.} \label{fig6}
\end{figure}

\bigskip
{\bf Antenna performance}

That the fiber-coupled THz antennas perform well at low
temperatures and at high magnetic fields is shown in Fig. 6. Here,
an emitter and receiver, each coupled by 20 meters of optical
fiber, are mounted on the cryogenic probe and loaded in the
variable-temperature insert of an 18 T superconducting magnet. The
emitter and receiver are driven by 1.5 mW and 2.4 mW of average
laser power, respectively.  The receiver current is amplified
(10$^7$ V/A) and sent directly to a digitizing oscilloscope. The
amplitude of the measured THz signal varies only slightly upon
cooldown from 300K to 1.6K, exhibiting a weak maximum near 100K.
\begin{figure}
\epsfig{file=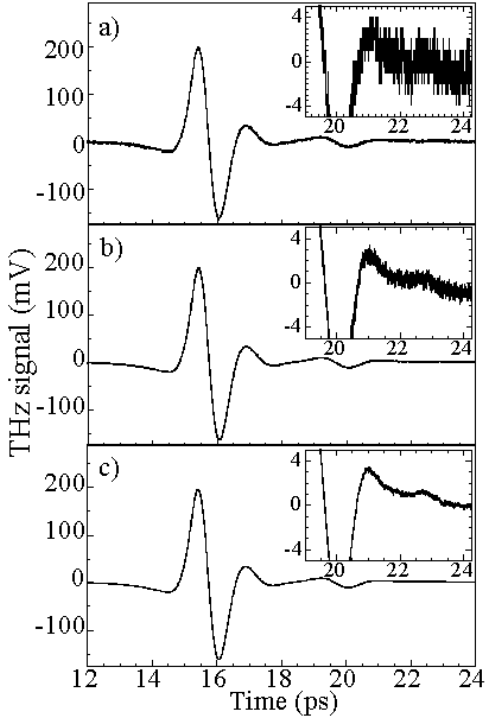, width=2.5in,clip=}
\vspace{0.1in} \caption{THz traces acquired in a) 25 ms, b) 100
ms, and c) 1600 ms. Insets show the noise level.} \label{fig7}
\end{figure}
The bandwidth of the THz radiation is unchanged by temperature.
THz emitters fabricated on standard semi-insulating GaAs
substrates are found to exhibit a much more pronounced temperature
dependence, in accord with recent studies by
Markelz\cite{Markelz}, where the changes are ascribed to the
strongly temperature-dependent mobility of electrons in GaAs. The
arrival times of the emitter and receiver optical pulses do vary
slightly with temperature, presumably due to unequal lengths of
thermally-contracted optical fiber in the cryogenic environment,
and this causes a small temporal (phase) shift in the measured THz
electric field, as can be seen in Fig. 6a at 90K and 2.2K. Because
the THz amplitude and phase do vary slightly with temperature, the
use of back-to-back sample and reference scans at each new
temperature (and magnetic field) is necessary to extract
quantitative conductivity data. Fig. 6b shows the performance of
the emitter/receiver pair (now immersed in 1.6K superfluid helium)
as a function of magnetic field up to 17 T. The amplitude of the
THz electric field is attenuated with field, although with no
change in bandwidth.  Most likely, the applied field acts to bend
the trajectories of the photoexcited electrons away from the axis
of the antenna dipoles, causing a reduction in signal. It is not
knows whether the field predominantly affects the efficiency of
the emitter, or the sensitivity of the receiver.

\begin{figure}
\epsfig{file=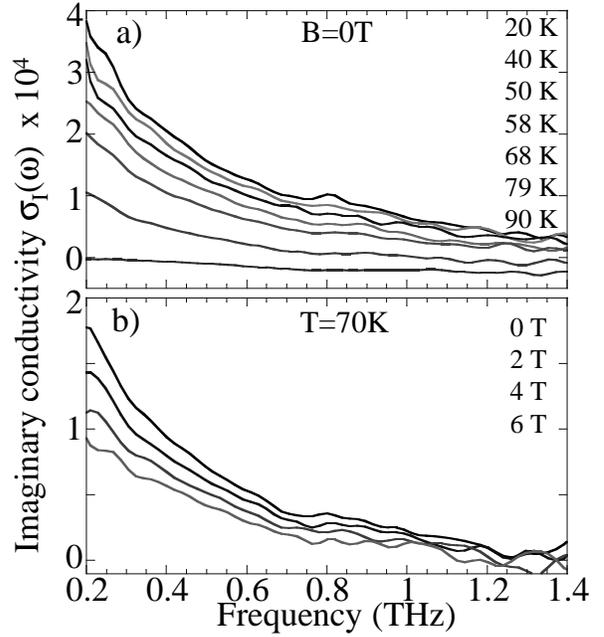, width=3.1in,clip=}
\vspace{0.1in} \caption{a) The imaginary conductivity of a 50 nm
thick YBCO film (T$_c$=85K) as a function of temperature, showing
suppression of phase-coherent superconductivity at high
temperatures. b) Same, except that here the superconductivity is
suppressed by application of magnetic field at 70K.  Sample
courtesy of Q. Jia (LANL).} \label{fig8}
\end{figure}
In the data of Fig. 6, 128 sweeps of the 20 Hz scanning
retro-reflector are digitally averaged by the oscilloscope, so as
to increase the signal-to-noise ratio. Since a long-term goal of
this project is to perform ultrafast coherent THz spectroscopy in
very high pulsed magnetic fields (and in particular during the 100
ms flat-top of the 60 Tesla Long-Pulse magnet at the Los Alamos
magnet lab\cite{Jaime,Crooker}), it is necessary that these
fiber-coupled devices exhibit sufficient signal-to-noise to
facilitate THz spectroscopy on short time scales.  Figure 7 shows
complete THz traces representing 25 ms (one half-cycle of the
retroreflector), 100 ms, and 1600 ms of acquired data.  The
measured root-mean-square noise on each scan (see insets) is 1.06
mV, 0.41 mV, and 0.12 mV, respectively, for signal-to-noise ratios
of approximately 190, 490, and 1700. Thus rather precise
spectroscopic THz measurements may be performed in high-field
pulsed magnets. Of course, in dc magnets (superconducting to 20 T,
or resistive to 45 T), no limitations on the amount of averaging
time are imposed, and the accuracy of the measurements may be
correspondingly increased.

\bigskip
{\bf Data}

To demonstrate the utility of the fiber-coupled THz antennas,
preliminary results on two material systems are shown in Fig. 8
and 9.  Here, the experimental probe is loaded into a cryogenic
vacuum can in the bore of a 7 T superconducting solenoid. Figure 8
shows the imaginary part of the measured THz conductivity of a
high-temperature superconducting YBCO film (50 nm thick, ${T_c
\sim}$85K).   At low temperatures (Fig. 8a), the ${1/\omega}$
conductivity from the Drude-like response of superconducting
particles with infinite scattering time is clearly observed. With
increasing temperature above ${T_c}$ this conductivity falls
rapidly, indicating the disappearance of phase coherent
superconductivity, in agreement with previous
works\cite{Nuss2,Averitt}.  Fig. 8b shows that similar behavior is
observed as a function of magnetic field for temperatures below
${T_c}$, indicating again that superconductivity is being
suppressed, but this time by the application of magnetic field
rather than temperature. By using applied magnetic fields to
suppress the superconducting state, these studies will permit
investigation of the terahertz complex conductivity of the
interesting {\it normal} state of high-${T_c}$ superconductors at
low temperatures {\it below} the zero-field ${T_c}$, where
transport (zero frequency) measurements in pulsed fields have
yielded a rich behavior\cite{Boebinger}.

\begin{figure}
\epsfig{file=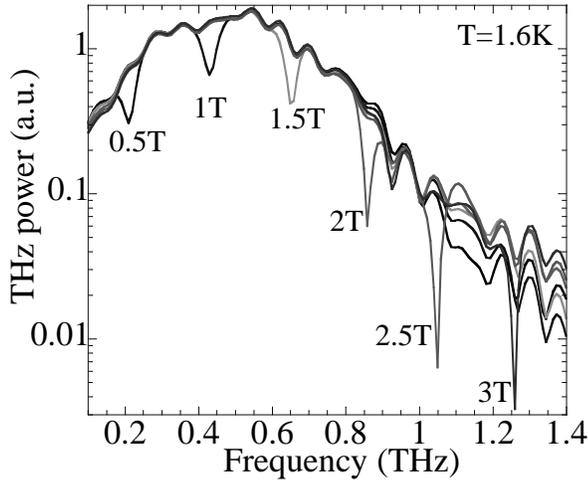, width=3.1in,clip=}
\vspace{0.2in} \caption{Low-field
dependence of the power spectrum of the raw time-domain THz data
upon passage through a high-mobility 2D electron gas at 1.6K.
Strong absorption resonances correspond to cyclotron motion.
Sample courtesy of M. Lilly (Sandia National Laboratory).}
\label{fig9}
\end{figure}

Lastly, Fig. 9 shows data on a very different system; namely, a
very high mobility 2-dimensional electron gas (${\mu=10^7}$
cm$^2$/V-s) formed at a GaAs/AlGaAs heterojunction.  Here, the raw
power spectrum of the transmitted THz pulse is shown in the
low-field regime where the electron cyclotron energy falls within
the THz detection bandwidth.  Clear oscillations in the
time-domain data (not shown) correspond to the observed cyclotron
absorption resonance, which evolves with the expected behavior
(${\hbar\omega_c=eB/m^*c}$ = 1.73 meV/T  = 420 GHz/Tesla). The
additional oscillations in the power spectrum are an artifact
arising from a multiple reflection of the THz pulse which appears
${\sim}$12 ps later in the time domain, and which may be avoided
by stacking additional ``dummy" wafers of GaAs onto the back of
the 2DEG sample. Combined with a carrier density modulation
scheme\cite{Some} that is synchronized with the scanning
retroreflector and digitizer, very sensitive density- and
field-dependent studies of the THz conductivity of ultraclean 2D
electron systems in the fractional quantum Hall regime may be
performed, providing deeper insight into the dynamics and
interactions of composite fermions (although generally such
studies require millikelvin temperatures and the present apparatus
would require adaptation to a dilution refrigerator.)

S.A.C. gratefully acknowledges A. J. Taylor for involvement in
this project, and M. Hanson and A. C. Gossard (UCSB) for the
ErAs/GaAs material. This work was supported by the NHMFL In-House
Research Program.


\begin{references}

\bibitem{Smith}
P. R. Smith, D. H. Auston, and M. C. Nuss, IEEE J. Quant. Elect.
{\bf QE-24}, 255 (1988).

\bibitem{Fattinger}
C. Fattinger and D. Grischkowsky, Appl. Phys. Lett. {\bf 54}, 490
(1989); M. van Exeter, C. Fattinger, and D. Grischkowsky, Appl.
Phys. Lett. {\bf 55}, 337 (1989); N. Katzenellenbogen and D.
Grischkowsky, Appl. Phys. Lett. {\bf 58}, 222 (1991).

\bibitem{Grisch}
D. Grischkowsky, S. Keiding, M. van Exeter, and C. Fattinger, J.
Opt. Soc. Am. B {\bf 7}, 2006 (1990).

\bibitem{Nuss1}
M. C. Nuss et al., J. Appl. Phys. {\bf 70}, 2238 (1991);

\bibitem{Nuss2}
See, e.g., M. C. Nuss {\it et al.}, Appl. Phys. Lett. {\bf 58},
2561 (1991); A. Frenkel {\it et al.}, Phys. Rev. B {\bf 54}, 1355
(1996); J. Corson {\it et al.}, Nature {\bf 398}, 221 (1999).

\bibitem{Pedersen}
J. E. Pedersen and S. R. Keiding, IEEE J. Quant. Elect. {\bf 28},
2518 (1992); D. M. Mittleman {\it et al.}, Chem. Phys. Lett. {\bf
275}, 332 (1997).

\bibitem{Cheville}
R. A. Cheville and D. Grischkowsky, Optics Lett. {\bf 20}, 1646
(1995).

\bibitem{Harde}
H. Harde, N. Katzenellenbogen, and D. Grischkowsky, J. Opt. Soc.
Am. B, {\bf 11}, 1018 (1994). D. M. Mittleman {\it et al.}, Appl.
Phys. B, {\bf B67}, 379 (1998).

\bibitem{Boebinger}
G. S. Boebinger {\it et al.}, Phys. Rev. Lett. {\bf 77}, 5417
(1996); S. Ono {\it et al.}, Phys. Rev. Lett. {\bf 85}, 638
(2000).

\bibitem{Jaime}
M. Jaime {\it et al.}, Nature {\bf 405}, 160 (2000); G. S.
Boebinger, A. Passner, P.C. Canfield, Z. Fisk, Physica B {\bf 21},
227 (1995).

\bibitem{Tokura}
Y. Tokura and N. Nagaosa, Science {\bf 288}, 462 (2000).

\bibitem{Uji}
S. Uji {\it et al.}, Nature {\bf 410}, 908 (2001); L. Balicas
{\it et al.}, Phys. Rev. Lett. {\bf 87}, 067002 (2001).

\bibitem{Burke}
P. J. Burke, J. P. Eisenstein, L. N. Pfeiffer, K. W. West, Rev.
Sci. Inst. {\bf 73}, 130 (2002).

\bibitem{Shen}
S. Shen and A. M. Weiner, IEEE Photon. Tech. Lett. {\bf 11}, 827
(1999); C. C. Chang, H. P. Sardesai and A. Weiner, Optics Lett.
{\bf 23}, 283 (1998).

\bibitem{Washburn}
B. R. Washburn, J. A. Buck, and S. E. Ralph, Optics Lett., {\bf
25}, 445 (2000).

\bibitem{Clark}
S. W. Clark, F. Ilday, and F. W. Wise, Optics Lett. {\bf 26},
1320 (2001).

\bibitem{Brener}
I. Brener {\it et al.}, Optics Lett. {\bf 21}, 1924 (1996); Y.
Cai {\it et al.}, Appl. Phys. Lett. {\bf 71}, 2076 (1997).

\bibitem{Kadow}
C. Kadow {\it et al.}, Appl. Phys. Lett. {\bf 75}, 3548 (1999); C.
Kadow {\it et al.}, Physica E {\bf 7}, 97 (2000); C. Kadow {\it et
al.}, Appl. Phys. Lett. {\bf 76}, 3510 (2000).

\bibitem{Rudd}
www.picometrix.com; see also J. V. Rudd, D. Zimdars, M.
Warmuth, Proc. of the SPIE {\bf 3934}, 27 (2000).

\bibitem{Markelz}
A. Markelz and E. J. Heilweil, Appl. Phys. Lett. {\bf 72}, 2229
(1998).

\bibitem{Crooker}
S. A. Crooker {\it et al.}, Phys. Rev. B {\bf 60}, R2173
(1999).

\bibitem{Averitt}
R. D. Averitt {\it et al.}, Phys. Rev. B {\bf 63}, 140502
(2001).

\bibitem{Some}
D. Some and A. Nurmikko, Appl. Phys. Lett. {\bf 65}, 3377
(1994).

\end{references}
\end{document}